\documentclass[manuscript]{aastex}

\slugcomment{Not to appear in Nonlearned J., 45.}
\usepackage{amsmath}
\DeclareMathOperator\arccosh{arccosh}

\shorttitle{Unstable Rossby waves in the Solar Tachocline}
\shortauthors{Djorgovski et al.}

\begin{document}

\title{Magneto-Rossby waves in the solar tachocline and the annual variations in solar activity}

    \author{Tamar Gachechiladze\altaffilmark{1}, Teimuraz V. Zaqarashvili\altaffilmark{2,1,3}, Eka Gurgenashvili\altaffilmark{1,4}, Giorgi Ramishvili\altaffilmark{1}, Marc Carbonell\altaffilmark{5,7}, Ramon Oliver\altaffilmark{6,7}, and Jose Luis Ballester\altaffilmark{6,7}}

\altaffiltext{1}{Abastumani Astrophysical Observatory at Ilia State university, Tbilisi, Georgia}
\altaffiltext{2}{Space Research Institute of Austrian Academy of Sciences, Graz, Austria}
\altaffiltext{3}{Institute of Physics, IGAM, University of Graz, Universit\"atsplatz 5, 8010, Graz, Austria}
\altaffiltext{4}{University of G\"ottingen, Germany}
\altaffiltext{5}{Departament de Matematiques i Informatica, Universitat de les Illes Balears, E-07122 Palma de Mallorca, Spain}
\altaffiltext{6}{Departament de Fisica, Universitat de les Illes Balears, E-07122 Palma de Mallorca, Spain}
\altaffiltext{7}{Institut d'Aplicacions Computacionals de Codi Comunitari (IAC3), Universitat de les Illes Balears,Spain}

\begin{abstract}
Annual oscillations have been detected in many indices of solar activity during many cycles. Recent multi spacecraft observations of coronal bright points revealed slow retrograde toroidal  phase drift (with the speed of $\sim$ 3 m s$^{-1}$) of 1 yr oscillations, which naturally suggested their connection with Rossby-type waves in the interior. We have studied from a theoretical point of view the dynamics of global magneto-Kelvin and magneto-Rossby waves in the solar tachocline with toroidal magnetic field. Using spherical coordinates, the dispersion relations of the waves and latitudinal structure of solutions were obtained analytically. We have also obtained the spectrum of unstable magneto-Rossby wave harmonics in the presence of the latitudinal differential rotation. Estimated periods and phase speeds show that the magneto-Rossby waves rather than the Kelvin waves match with the observations of 1 yr oscillations. On the other hand, Morlet wavelet analysis of Greenwich Royal Observatory sunspot areas for the solar cycle 23 has revealed multiple periodicities with  periods of 450-460 days, 370-380 days, 310-320 days, 240-270 days, and 150-175 days in hemispheric and full disk data. Comparison of theoretical results with the observations allow us to conclude that the global magneto-Kelvin waves in the upper overshoot tachocline may be responsible for the periodicity of 450-460 days ($\sim$ 1.3 yrs), while the remaining periods can be connected with different harmonics of global  fast magneto-Rossby waves.

\end{abstract}

\keywords{Sun: oscillations  ---
Sun: interior --- Sun: magnetic fields}

\section{Introduction}

Solar activity variation mainly occurs over  time scale of 11 years, which is called the solar cycle or Schwabe cycle \citep{Schwabe1844}. However, longer and shorter periodicities than the solar cycles are also seen in many different activity indices. The longer time scales are hundreds of years \citep{Gleissberg1939,Suess1980,solanki2013,Zaqarashvili2015}, but the shorter time scales are several months (155-160 days known as Rieger periodicity) and 1-2 years. The periodicity of 154 days was discovered \citep{Rieger84} in $\gamma$-ray flares and then it has been found in many indices of solar activity \citep{Lean89,carball90,oliver98}. Oscillations with the period of $\sim$ 2 years modulate almost all indices of solar activity \citep{sakurai81,gigolashvili95,knaack05,sykora10,laurenza12,vecchio12,Bazilevskaya2014,Kiss2018} and have been also detected by helioseismology \citep{Broomhall2012,simoniello2013}. These so called quasi biennial oscillations may be explained either by a double dynamo model \citep{Benevolenskaya1998} operating in two dynamo layers (one below the convection zone and another near the surface) or by magneto-Rossby wave instability in the solar tachocline \citep{Zaqarashvili2010b}. On the other hand, 1 yr oscillations (with period of 323 days) have been found in the sunspot blocking function, 10.7 cm radio flux, sunspot number, and plage index daily data during cycles 19-21 \citep{Lean89} and also in sunspot number and areas \citep{oliver92}. Recently, \citet{McIntosh2015,McIntosh2017} showed the existence of 1 yr periodicity in coronal bright points using simultaneous observations from Solar Dynamics Observatory (SDO) and STEREO. \citet{McIntosh2017} found that the waves moved with   a slow retrograde (i.e. opposite to the solar rotation) phase speed ($\sim$ 3 m s$^{-1}$) in the toroidal direction, which indicates  their connection with Rossby-type waves.

Rossby waves govern large-scale flow dynamics on a rotating sphere and owe
their existence to the latitudinal variation of Coriolis parameter. The waves are well studied in the Earth's atmosphere and oceans \citep{Rossby1939,lon65,gill82,ped87,Matsuno1966}. They may also play an important role in the dynamics of the solar interior/atmosphere \citep{Gilman1969a,Gilman1969b,Sturrock1999,Sturrock2015,lou00,Zaqarashvili2010a,Zaqarashvili2010b,Zaqarashvili2015,dikpati2017,dikpati2018}. Recent observations of Rossby waves in coronal bright points \citep{McIntosh2017} and in subsurface velocity \citep{Loptien2018} greatly increased interest in these waves. The waves may play an especially important role in the tachocline, which is a thin layer between  the differentially rotating convection zone and the solidly rotating radiative envelope \citep{spi92}. The sub-adiabatic temperature gradient leads to the significant decrease of the effective gravity in the upper part of the tachocline \citep{gil00,dik05}, which then causes the trapping of shallow water waves around the equator \citep{Zaqarashvili2018}. The equatorial trapping of shallow water waves is well known in fluids on rapidly rotating spheres like the Earth \citep{lon65,longuet68,Matsuno1966,bouchut2005}. The equatorial waves may lead to the observed Rieger-type and annual oscillations \citep{lou00,Zaqarashvili2018a}.

The connection of 1 yr periodicity with the solar tachocline currently is not clearly defined. It is very important that a similar periodicity of 1.3 yrs was found  in the rotation of tachocline during the cycle 23 by helioseismology \citep{howe00}. \citet{Richardson94} using data obtained by IMP-8 and Voyager 2 spacecrafts found a very strong modulation in the solar wind speed with a period of 1.3 years.  \citet{Krivo02}  analyzed different sunspot data sets and reported that the power at the 1.3-year periodicity fluctuates considerably with time, being stronger during stronger sunspot cycles. On the other hand, \citet{Gurgenashvili2016,Gurgenashvili2017} showed that the Rieger periodicity significantly depends on the strength of the cycles and level of hemispheric magnetic activity. Therefore, it is  very important to test if the periodicities of 1 yr and 1.3 yrs belong to the same oscillation branch. The two periodicities could be also accompanied by oscillations at other time scales and thus could be revealed by wavelet or Fourier analyses of the same data. The oscillations might be connected with the dynamo layer in the solar interior, therefore sunspots or sunspot areas are most useful traces of the periodicity.

In this paper, we use  the shallow water magnetohydrodynamic (MHD) equations \citep{gil00} to study the dynamics of Rossby and Kelvin waves in the tachocline and their possible connection with observed mid-range periodicity (hundreds of days) in solar activity. We have also analysed Greenwich Royal Observatory hemispheric sunspot area data during solar cycle 23 in order to search for multiple periodicities in  the northern and southern hemispheres.

\section{Main equations}

 We use a spherical coordinate system $(r, \theta, \phi)$  rotating with the solar equator, where $r$ is the radial
coordinate, $\theta$  is the co-latitude and $\phi$ is the longitude.

The magnetic field is predominantly toroidal in the solar tachocline, therefore we use $\vec B = B_{\phi}(\theta) \sin \theta \hat{e}_{\phi}$, where $B_{\phi}$ is in general a function of co-latitude. We also consider the observed latitudinal differential rotation in the form  \citep{Howard1970}
\begin{equation}\label{omega}
\Omega=\Omega_0 (1 - s_2 \cos^2 \theta -s_4 \cos^4 \theta)=\Omega_0 + \Omega_1(\theta),
\end{equation}
where $\Omega_0$ is the equatorial angular velocity, and $s_2, s_4$ are constant parameters determined by observations.  These parameters are slightly smaller than  their photospheric values at the upper part of the tachocline, but tend to be very small near the bottom \citep{Schou1998}

Then, the linear shallow water MHD equations (Gilman 2000) can be rewritten in the frame  rotating (at $\Omega_0$) as follows:
\begin{equation} \label{MHD1}
{{\partial u_{\theta}}\over {\partial t}} +
\Omega_1{{\partial u_{\theta}}\over {\partial \phi}} - 2\Omega\cos \theta u_{\phi} = -{{g}\over
{R}}{{\partial h}\over {\partial \theta}}+ {{B_{\phi}}\over
{{4\pi\rho R}}}{{\partial b_{\theta}}\over {\partial \phi}}-2
{{B_{\phi} \cos \theta}\over {{4\pi\rho R}}}b_{\phi},
\end{equation}
\begin{equation}\label{MHD2}
{{\partial u_{\phi}}\over {\partial t}} +
\Omega_1{{\partial u_{\phi}}\over {\partial \phi}}  +\frac{u_{\theta}}{\sin{\theta}}{{\partial [\Omega\sin^2{\theta}]}\over
{\partial \theta}}=-{{g}\over {R
\sin \theta}}{{\partial h}\over {\partial \phi}}+{{B_{\phi}}\over
{{4\pi\rho R}}}{{\partial b_{\phi}}\over {\partial \phi}}+
{{b_{\theta}}\over {{4\pi\rho R \sin \theta}}}{{\partial [B_{\phi}\sin^2 \theta]}\over
{\partial \theta}},
\end{equation}
\begin{equation}\label{MHD3}
{{\partial b_{\theta}}\over {\partial t}} +
\Omega_1{{\partial b_{\theta}}\over {\partial \phi}} = {{B_{\phi}}\over {{R}}}{{\partial u_{\theta}}\over
{\partial \phi}},
\end{equation}
\begin{equation}\label{MHD4}
{{\partial }\over {\partial \theta}}\left (\sin \theta b_{\theta}
\right ) + {{\partial b_{\phi}}\over {\partial
\phi}}+\frac{B_{\phi}\sin \theta}{H}{{\partial h}\over {\partial
\phi}}=0,
\end{equation}
\begin{equation}\label{MHD5}
{{\partial h}\over {\partial t}} +
\Omega_1{{\partial h}\over {\partial \phi}}
+\frac{H}{R\sin{\theta}}{{\partial }\over {\partial
\theta}}\left (\sin \theta u_{\theta} \right ) + \frac{H}{R
\sin{\theta}}{{\partial u_{\phi}}\over {\partial \phi}}=0,
\end{equation}
where $u_{\theta}$, $u_{\phi}$, $b_{\theta}$ and $b_{\phi}$ are the
velocity and magnetic field perturbations, $H$ is the tachocline
thickness and $h$ is its perturbation, $\rho$ is the density, $g$ is the gravity and
$R$ is the distance from the solar center to the tachocline.
These equations contain magneto-Rossby and magneto-gravity waves. The important parameter governing the shallow water system is
\begin{equation}\label{omega}
\epsilon={{\Omega^2_0R^2}\over {gH}},
\end{equation}
where $\sqrt{gH}$ is the surface gravity speed. When $\epsilon \ll 1$ (i.e. the case of slow rotation), then the Rossby waves are decoupled from the gravity waves and they can be considered on a spherical surface. On the other hand,  when $\epsilon \gg 1$ (the case of fast rotation or reduced gravity), then the shallow water waves tend to be localised near the equator.
The sub-adiabatic temperature gradient in the upper overshoot part of the tachocline provides a negative buoyancy force to the deformed upper surface, and hence leads to a reduced gravity so that the surface feels less gravitational field \citep{gil00}.  \citet{Dikpati2001} showed that the dimensionless value of reduced gravity $G=1/\epsilon=gH/(R^2 \Omega^2_0)$ is in the range of $G>100$ in the radiative part of the tachocline and in the range of $10^{-3} \leq G \leq 10^{-1}$ in the upper overshoot part. Recently \citet{Zaqarashvili2018}  showed that the equatorial magneto-Kelvin waves lead to annual oscillations in the case of reasonable values of reduced gravity (in rectangular geometry).  The dispersion relation of Kelvin waves is  $\omega/\Omega_0 \sim \sqrt{G} k R$, where $\omega$ and $k$ are the frequency and the  wavenumber of the waves. Then for the large-scale waves with $k R \approx 1$, 1 yr periodicity is achieved when $G\approx5$ 10$^{-3}$, which corresponds to the upper tachocline. Therefore, the Kelvin waves might lead to 1 yr periodicity when $G\ll 1$ or $\epsilon \gg 1$. On the other hand, the annual oscillations can be also achieved owing to Rossby waves on a spherical surface, which obey the dispersion relation $\omega=-2m\Omega_0/n(n+1)$, where $m$ ($n$) is toroidal (poloidal) wavenumber (the dispersion relation is valid for $\epsilon \ll 1$). For  $m=1$, one may get $\sim$ 1 yr periodicity for $n=5$ or $n=6$. If $\epsilon \gg 1$ then the period of Rossby waves increases up to the length of solar cycle \citep{Zaqarashvili2018}. Consequently, the Rossby wave might give 1 yr periodicity in the lower part of tachocline with $\epsilon \ll 1$. Therefore, in the following we consider the magneto-Kelvin waves in the upper part of the tachocline (with $\epsilon \gg 1$) and the magneto-Rossby waves in the lower part of the tachocline (with $\epsilon \ll 1$). We study  both cases separately.

In the following we use typical tachocline parameters such as $\Omega_0$=2.6 10$^{-6}$ s$^{-1}$, $\rho$=0.2 g cm$^{-3}$, and
$R$=5 10$^{10}$ cm.

\section{Spherical equatorial magneto-Kelvin waves for $\epsilon \gg 1$}

Here we consider the upper tachocline, where $\epsilon \gg 1$, with rigid body rotation ($\Omega_1=0$). Kelvin waves have zero poleward velocity in rectangular coordinates (Matsuno 1966). On the other hand, the poleward component of the waves is not exactly zero in spherical coordinates, but it is significantly smaller than the toroidal component (Longuet-Higgins 1968). Therefore, in order to derive the dispersion relation of equatorial magneto-Kelvin waves one can easily set up negligible values for poleward velocity and magnetic field components, $u_{\theta}=b_{\theta}=0$. Then Eqs.~(\ref{MHD1})-(\ref{MHD5})  can be rewritten after Fourier transform with $\exp(-i\omega t + i m \phi)$ as
\begin{equation} \label{kelvin1}
 2\Omega_0 \cos \theta u_{\phi} = {{g}\over
{R}}{{\partial h}\over {\partial \theta}}+2{{B_{\phi} \cos \theta}\over {{4\pi\rho R}}}b_{\phi},
\end{equation}
\begin{equation}\label{kelvin2}
\omega u_{\phi} ={{m g}\over {R\sin \theta}} h-{{m B_{\phi}}\over {{4\pi\rho R}}}b_{\phi},
\end{equation}
\begin{equation}\label{kelvin3}
b_{\phi}=-\frac{B_{\phi}\sin \theta}{H} h,
\end{equation}
\begin{equation}\label{kelvin4}
\omega h = m \frac{H}{R
\sin{\theta}}u_{\phi}.
\end{equation}
Eqs.~(\ref{kelvin2})-(\ref{kelvin4}) lead to the dispersion equation
\begin{equation} \label{kelvindisp}
 \omega^2-{{m^2}\over {R^2}}\left ( {{c^2}\over { \sin^2 \theta}} +V^2_A  \right )=0,
\end{equation}
where $c=\sqrt{gH}$ is the surface gravity speed and $V_A=B_{\phi}/\sqrt{4\pi \rho}$ is the Alfv\'en speed. Near the equator $\sin^2 \theta \approx 1$ and we have
\begin{equation} \label{kelvineq}
\left (\omega-{{m}\over {R}}\sqrt{c^2 +V^2_A}\right )\left ( \omega+{{m}\over {R}}\sqrt{c^2 +V^2_A}\right )=0.
\end{equation}
The dispersion equation has two solutions. One should check which of the two solutions satisfies the bounded boundary condition with latitude.
We will consider two different profiles of the toroidal magnetic field.

\begin{figure}
   \begin{center}
  \includegraphics[width=15.0cm]{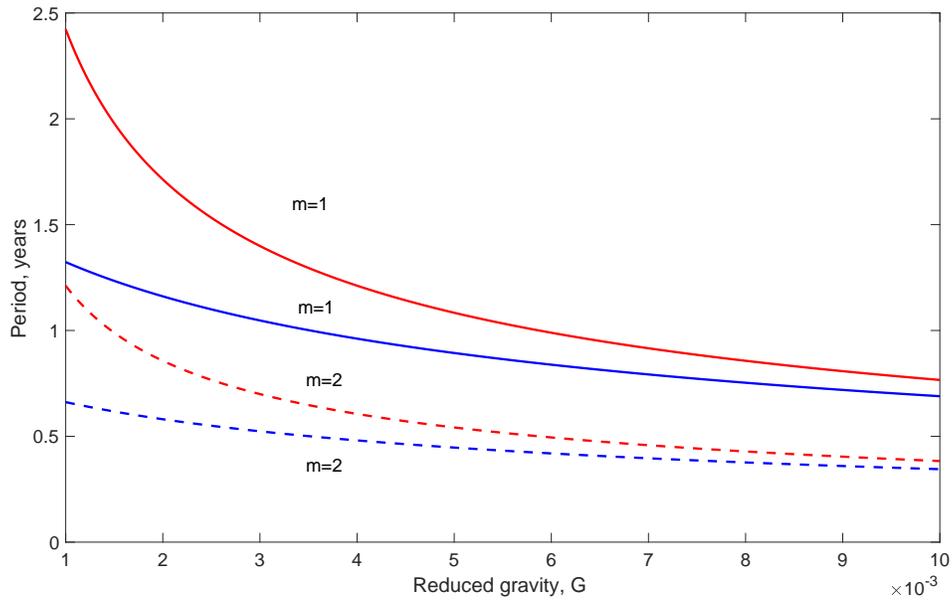}
  \end {center}
  \caption{Period of equatorial Kelvin waves vs normalised reduced gravity, $G$. Thick (dashed) red line shows the solution for $m=1$ ($m=2$) harmonic in the case of the nonuniform magnetic field.
Thick (dashed) blue line shows the solution for $m=1$ ($m=2$) harmonic in the case of the uniform magnetic field with strength of 10 kG.}\label{Kelvin}
\end{figure}

\subsection{Uniform magnetic field}

First we consider the uniform magnetic field $B_{\phi}=B_0=const$. Then Eq.~(\ref{kelvin1}) leads to the solution near the equator in the form
\begin{equation} \label{kelvinuni1}
h=h_0  \exp{\left (-{{\Omega R^2 \omega +mV^2_A}\over {m c^2}} {\hat \theta}^2\right )},
\end{equation}
where ${\hat \theta}=90^0-{\theta}$ is
the latitude. Therefore, only the positive frequency
\begin{equation} \label{kelvinuni2}
\omega={{m}\over {R}}\sqrt{c^2 +V^2_A}
\end{equation}
leads to the exponentially decaying solution with latitude  for relatively weak magnetic field
\begin{equation} \label{kelvinuni3}
h=h_0  \exp{\left (-{{\Omega R \sqrt{c^2 +V^2_A} +V^2_A}\over {c^2}} {\hat \theta}^2\right )}.
\end{equation}

\begin{figure}
   \begin{center}
  \includegraphics[width=15.0cm]{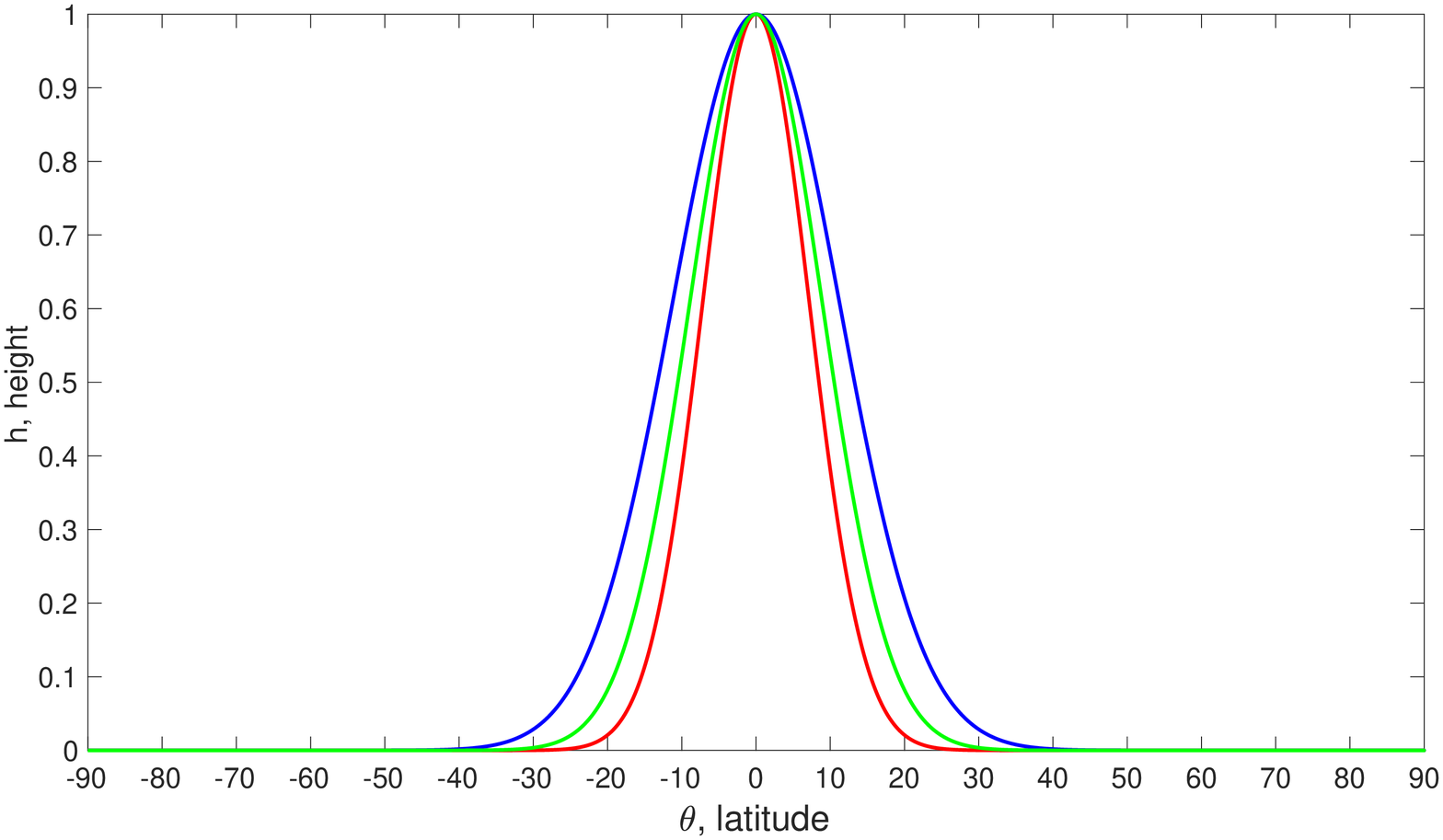}
  \end {center}
  \caption{Perturbation of layer height vs solar latitude in equatorial Kelvin waves with $m=1$.  Thick red (blue) line shows the solution for nonuniform magnetic field with the strength of 10 kG in the case of $G=0.001$ ($G=0.006$). Thick green line shows the solution for uniform magnetic field with the strength of 10 kG in the case of $G=0.004$.}\label{Kelvin}
\end{figure}

Figure 1 (blue lines) shows the period of equatorial Kelvin waves vs normalised reduced gravity, $G$, for the magnetic field strength of 10 kG from Eq.~(\ref{kelvinuni2}).
It is seen that the oscillations with period of $\sim$ 1 yr can be formed for $m=1$ when the normalised reduced gravity   has a value in the range of 0.003-0.006 (the period is exactly one year for $G= 0.004$). These values are in the range of normalised reduced gravity in the upper overshoot region of the tachocline as estimated by \citet{Dikpati2001}. For the same values of the reduced gravity, the harmonic with $m=2$ gives the Rieger-type periodicity. Stronger magnetic field will require much smaller reduced gravity.  The green line on Figure 2 displays the solution corresponding to Eq.~(\ref{kelvinuni3}). The solution is concentrated around the equator  as  expected from Eq.~(\ref{kelvinuni3}).

The dispersion relation, Eq.~(\ref{kelvinuni2}), was recently obtained by \citet{Marquez2017}, who showed that the solution with the negative frequency can be also equatorially trapped for strong magnetic field. This is also seen from Eq.~(\ref{kelvinuni1}), where the negative frequency solution can be also equatorially trapped if the second term in the exponent is larger than the absolute value of the first one. For small reduced gravity, this may happen when the magnetic field strength exceeds 200 kG, which is stronger than expected in the solar tachocline.

\subsection{Nonuniform magnetic field}

In the second case, a nonuniform magnetic field with the profile $B_{\phi}=B_0\cos \theta \approx B_0 {\hat \theta}$ \citep{gilman97}  is considered, which leads to the bounded solution
\begin{equation} \label{kelvin31}
h=h_0  \exp{\left (-{{\Omega R}\over {c}} {\hat \theta}^2 -{{V^2_A}\over {2c^2}} {\hat \theta}^4 \right )},
\end{equation}
with the dispersion relation
\begin{equation} \label{kelvin21}
\omega={{m c}\over {R}}.
\end{equation}

The dispersion relation, Eq.~(\ref{kelvin21}), shows that the period of Kelvin waves in the nonuniform magnetic field depends on the reduced gravity but not on the field strength. This result is straightforward as the magnetic  field is negligibly small around the equator owing to the  profile used. The oscillations with the periodicity of $\sim$ 1 yr (see Figure 1) occur when the normalised reduced gravity
 has a value in the range of 0.005-0.007 (the period is exactly one year for $G$=0.006). The  range of required normalised reduced gravity corresponds to the conditions of the upper  tachocline  \citep{Dikpati2001}.
At the same value of the reduced gravity, the $m=2$ mode gives the Rieger type periodicity. It should be also mentioned that the reduced gravity  of 0.0015 gives the annual oscillations for the $m=2$ mode. The solutions corresponding to m=1 harmonics are plotted on Figure 2, which show that they  are concentrated around the equator.  The solution with $m=1$ for $G=0.001$ with magnetic field strength of 10 kG penetrates up to the $\pm$ 20 latitude, while the solution for $G=0.006$ with the same field strength penetrates up to the $\pm$ 30 latitude. Hence, the nonuniform magnetic field gives the latitude of observed periodicity in the appearance of coronal bright points  \citep{McIntosh2015,McIntosh2017}. On the other hand, Eqs.~(\ref{kelvinuni2}) and (\ref{kelvin21}) show that the Kelvin waves are prograde i.e. they propagate in the direction of solar rotation, which is exactly opposite to the observation by  \citet{McIntosh2017}. The only exception is a very strong uniform magnetic field with a strength of $>$ 200 kG, which gives retrograde propagation of equatorial magneto-Kelvin waves (see previous subsection).

\section{Spherical magneto-Rossby waves on spherical surface for $\epsilon \ll 1$}

In the upper  tachocline with $\epsilon \gg 1$ the magneto-Rossby waves are concentrated near the equator and their characteristic period is increased up to the solar cycle time scale \citep{Zaqarashvili2018}. Therefore,  in order to get 1 yr oscillations, we consider the lower tachocline with the stable stratification i.e. $\epsilon \ll 1$.  In this case, we can use the 2D incompressible linearized equations in the frame rotating with $\Omega_0$, which can be derived from equations $(2)-(6)$
 \citep{Zaqarashvili2010a}
\begin{equation}\label{mg1}
 {{\partial u _\theta}\over {\partial t}}+{\Omega_1}{{\partial u _\theta}\over {\partial \phi}}-{2\Omega\cos \theta u_\phi}={-{{1\over {\rho R}}{{\partial
p_t} \over {\partial \theta}}+ {{{B_{\phi}}} \over {4 \pi \rho R}} {{\partial b_\theta} \over {\partial \phi}}-2 {{{B_{\phi}}\cos\theta} \over {4\pi \rho R}}b_\phi}},
\end{equation}
 \begin{equation}\label{mg2}
{{\partial u_\phi} \over {\partial t}}+ {\Omega_1} {{\partial u _\phi}\over {\partial \phi}} + {2\Omega\cos \theta u_\theta}+ {u_\theta}\sin\theta{{\partial \Omega_1}\over {\partial \theta}} = -{1\over {\rho R \sin\theta}}{{\partial p_t} \over {\partial \phi}} +{{B_{\phi}}\over {4\pi \rho R}}
{{\partial b_{\phi}}\over {\partial \phi}}
+{{b_{\theta}}\over {4\pi \rho R\sin{\theta}}} {{\partial [B_{\phi} \sin^2{\theta}]}\over {\partial \theta}},
\end{equation}
\begin{equation}\label{mg3}
{{\partial b_\theta} \over {\partial t}} + {\Omega_1 ({\theta})} {{\partial b _\theta}\over {\partial \phi}}=  {{{{B_{\phi}}}} \over { R}} {{\partial u_\theta} \over {\partial \phi}},
\end{equation}
\begin{equation}\label{mg4}
{{\partial [\sin{\theta}b_{\theta}]}\over {\partial \theta}}+
{{\partial b_{\phi}}\over {\partial \phi}}=0
\end{equation}
\begin{equation}\label{mg5}
 {{\partial [\sin{\theta}u_{\theta}]}\over {\partial \theta}}+
 {{\partial u_{\phi}}\over {\partial \phi}}=0,
\end{equation}
where $p_t$ is the total (thermal + magnetic) pressure.

Considering the stream functions for the velocity and the magnetic field,
 \begin{equation}\label{stream}
u_{\theta}={1\over {{\sin \theta}}}{{\partial \Psi}\over {\partial
\phi}},\,\,u_{\phi}=-{{\partial \Psi}\over {\partial \theta}},\,\,
b_{\theta}={1\over {{\sin \theta}}}{{\partial \Phi}\over {\partial
\phi}},\,\,b_{\phi}=-{{\partial \Phi}\over {\partial \theta}}
\end{equation}
and using a Fourier analysis of the form $\exp[i(m\phi-\omega{t})]$ one can arrive at following two equations \citep{Gurgenashvili2016}
\begin{equation}\label{B}
B {\Psi}=(\Omega_d-\omega_1 )\Phi,
\end{equation}
$$
(\Omega_d-\omega_1)
\left [{{\partial (1-\mu^2)}\over {\partial \mu}} {{\partial}\over {\partial \mu}}-{{m^2}\over {1-\mu^2}}\right ]{\Psi}+
\left [ 2- {{d^2}\over {d \mu^2}}[\Omega_d (1-\mu^2)] \right ]\Psi-
$$
\begin{equation}\label{Omega}
\beta^2 B\left [{{\partial(1-\mu^2)}\over {\partial \mu}}{{\partial }\over {\partial \mu}}-{{m^2}\over {1-\mu^2}}\right ]{\Phi}+
\beta^2{{\partial^2(B(1-\mu^2))}\over {\partial \mu^2}}{\Phi}=0,
\end{equation}
where $\mu=\cos\theta$, $\Psi$ is normalized by $\Omega_0R$, $\Phi$ is normalized by $B_0$, and
$$ \Omega_d(\mu)={\Omega_1(\mu)\over{\Omega_0}}, \omega_1={\omega\over{m\Omega_0}},$$
\begin{equation}\label{Parameters}
\beta^2={{B_0^2}\over {4\pi \rho \Omega_0^2 R^2}}, B(\mu)={{B_\phi (\mu)}\over {B_0}}.
\end{equation}

Solution of the equations is complicated owing to the existence of magnetic field and differential rotation. We will use two different approximations to find the solutions of magneto-Rossby waves.

\begin{figure}
   \begin{center}
  \includegraphics[width=18.0cm]{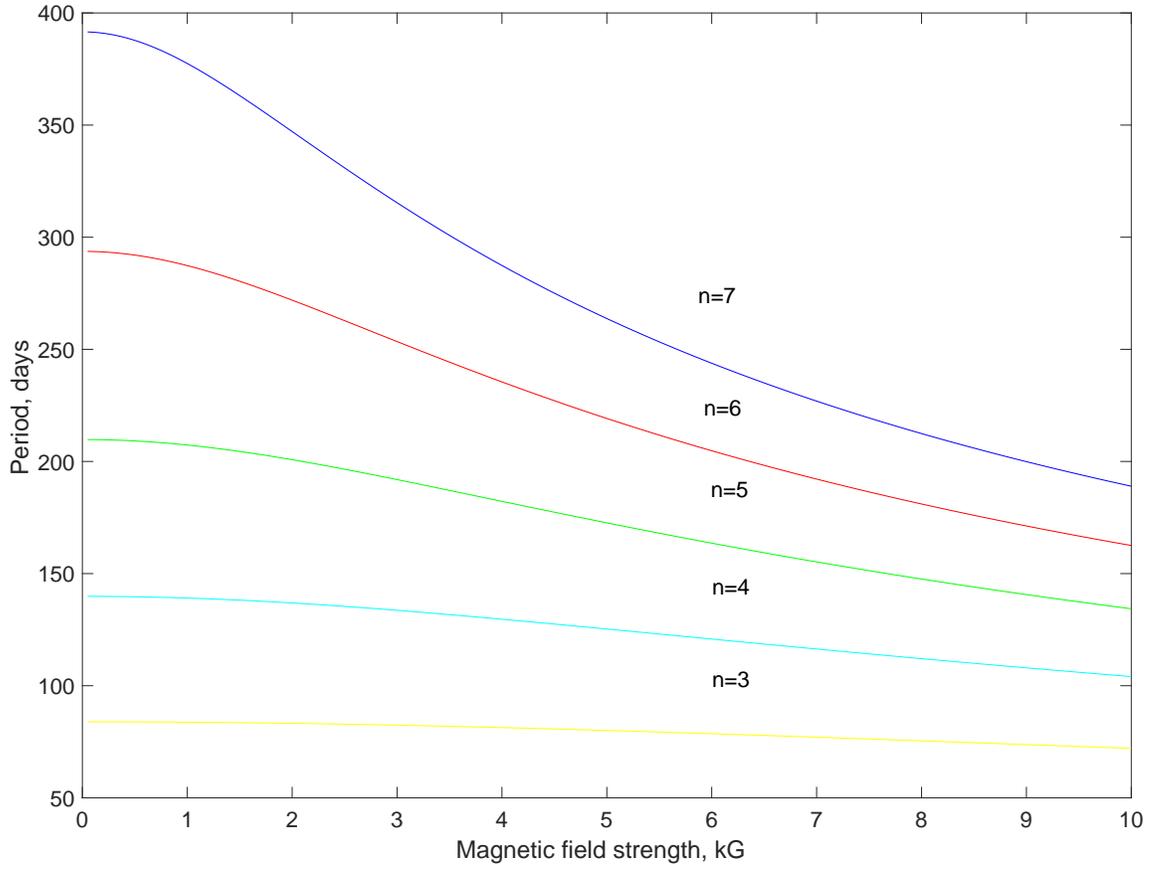}
  \end {center}
  \caption{Period of spherical magneto-Rossby wave harmonics with toroidal wavenumber $m=1$ vs maximal magnetic field strength, $B_0/2$.  Green, red, blue, and magenta lines correspond to $n=3$, $n=4$, $n=5$, and $n=6$  harmonics, respectively. }\label{Kelvin1}
\end{figure}

\subsection{Magneto-Rossby waves on rigidly rotating surface}

Let us first consider the approximation of rigid rotation, i.e. $\Omega_d=0$, which considerably simplifies the equations. If we use the magnetic field configuration in the form of $B_{\phi}=B_0\mu$, then the two equations can be cast in one equation:
\begin{equation}\label{rigid1}
{{\partial }\over {\partial \mu}}(1-\mu^2){{\partial H}\over {\partial \mu}}+\left [-{{m^2}\over {1-\mu^2}}  +{{\mu^2\beta^4}\over {(\omega_1^2-\beta^2\mu^2)^2}}(1-\mu^2)-{{2\omega_1-(1+7\mu^2)\beta^2}\over {\omega_1^2-\beta^2\mu^2}} \right ]H=0,
\end{equation}
where
\begin{equation}\label{rigid2}
H=-{{\Psi\sqrt{\omega_1^2-\beta^2\mu^2}}\over {\omega_1}}={{\Phi\sqrt{\omega_1^2-\beta^2\mu^2}}\over {\mu}}.
\end{equation}
 This equation has a singular point at $\omega_1=\beta \mu$, which could be a critical instability layer for some modes. As $\mu_{max}=1$, then the weak field approximation ($\beta \ll 1$) removes the singular point, which in the tachocline parameters requires a field strength $<$ 200 kG. Therefore, we consider the weak magnetic field approximation
\begin{equation}\label{ap}
\frac{\beta}{|\omega_1 |}\ll 1,
\end{equation}
  then Eq.~(\ref{rigid1}) is transformed into the equation (keeping second order terms with $\beta/\omega_1$)
\begin{equation}\label{rigid3}
{{\partial }\over {\partial \mu}}(1-\mu^2){{\partial H}\over {\partial \mu}}+\left [-{{m^2}\over {1-\mu^2}} +\lambda_{mn}+\gamma^2 \mu^2 \right ]H=0,
\end{equation}
where
\begin{equation}\label{rigid4}
\lambda_{mn}=-{{2\omega_1-\beta^2}\over {\omega_1^2}},  \gamma^2={{7\omega_1-2}\over {\omega_1}}{{\beta^2}\over {\omega_1^2}}.
\end{equation}
This is a spheroidal wave equation and has bounded solutions in terms of oblate spheroidal wave functions  $S_{mn}(\gamma,\mu)$. In the lowest order of $\gamma^2$,  eigenvalue
$\lambda_{mn}$ is expressed by the formula \citep{abramowitz}
\begin{equation}\label{rigid5}
\lambda_{mn}=n(n+1)+ {{1}\over {2}}  \left ( 1 - {{(2m-1)(2m+1)}\over {(2n-1)(2n+3)}} \right )\gamma^2.
\end{equation}
If we use again  the assumption $\beta/|\omega_1 |\ll 1$, then Eq.~(\ref{rigid4}) leads to $\gamma \approx 0$. Therefore, the last term with $\gamma^2$ can be ignored and one obtains the dispersion relation of magneto-Rossby waves  \citep{zaqa09}
\begin{equation}\label{rigid6}
n(n+1)\left ({{ \omega}\over {\Omega_0}}\right )^2+2m{{ \omega}\over {\Omega_0}}-m^2\beta^2=0.
\end{equation}

\begin{figure}
\includegraphics[width=17.0cm]{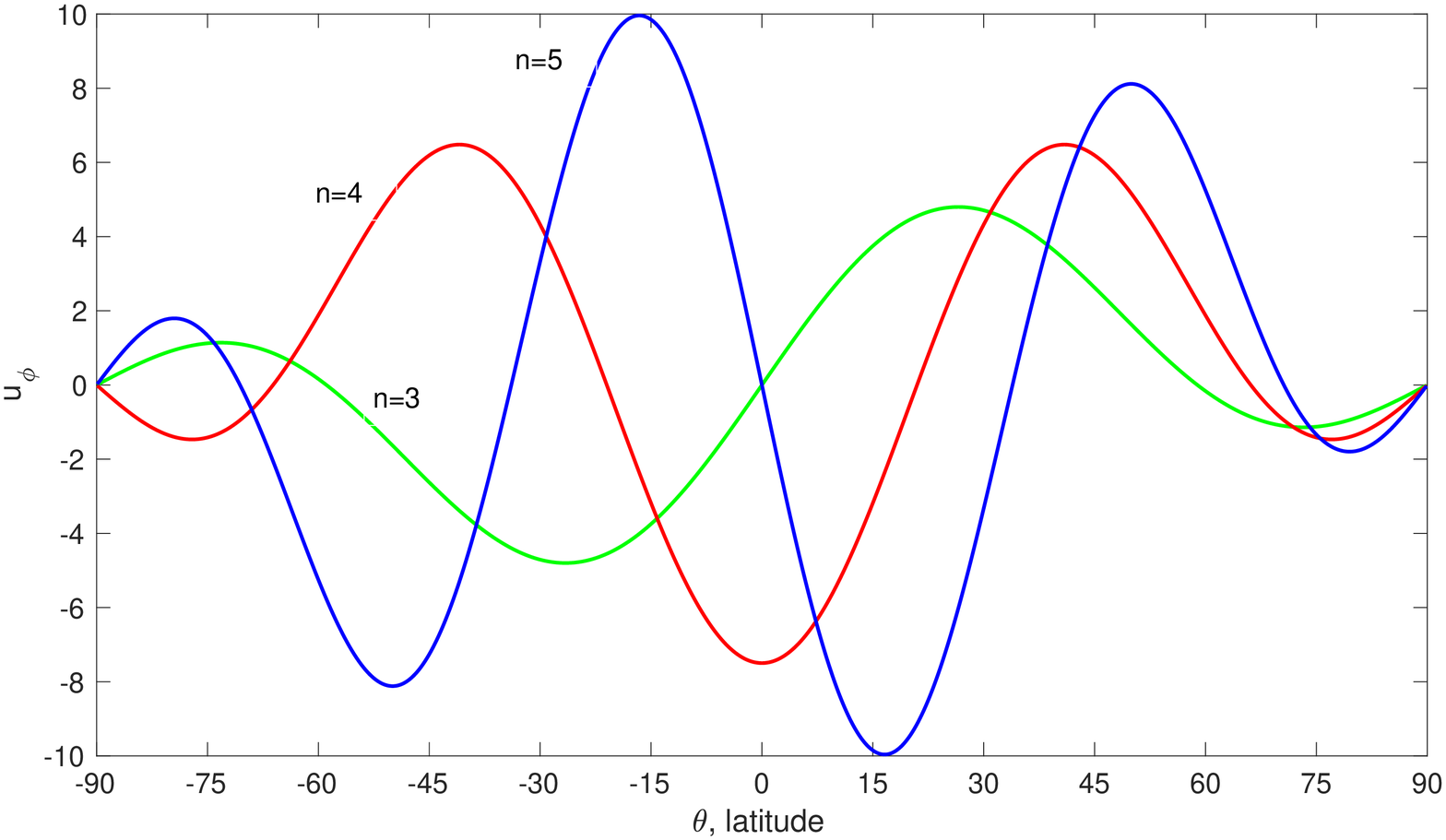}{\centering}
\includegraphics[width=17.0cm]{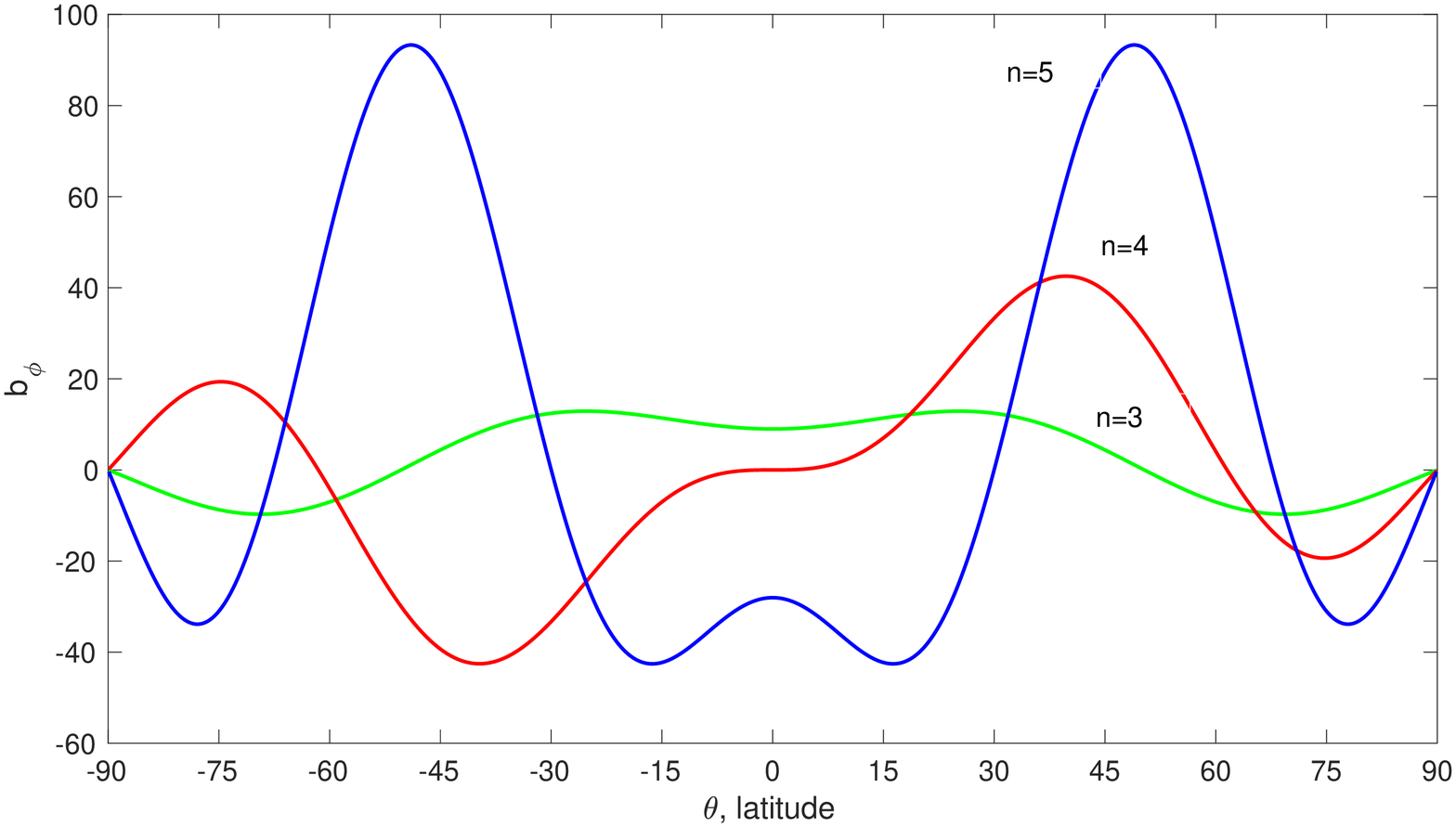}{\centering}
\caption{Toroidal velocity (upper panel) and magnetic field (lower panel)  perturbations, $u_{\phi}$ and $b_{\phi}$, vs latitude for the harmonics of $n=3$, $n=4$, and $n=5$ with $m=1$. The maximal value of unperturbed toroidal magnetic field strength is 1 kG.}
\end{figure}


The dispersion relation is a second order polynomial in frequency, therefore it describes two modes: fast and slow magneto-Rossby waves. The difference between the frequencies of the harmonics depends on the magnetic field strength. In the case of weak magnetic field strength, $\beta \ll 1$, the fast magneto-Rossby mode frequency is close to the hydrodynamic Rossby wave frequency i.e.
\begin{equation}\label{fast}
{{ \omega}\over {\Omega_0}} \approx - {{2m}\over {n(n+1)}}.
\end{equation}
In this case, ${{ \omega}/{\Omega_0}}$ is of the order of 0.1 for first several harmonics with $n$ and therefore $\gamma^2$ is smaller than 1 for the field strength of $<$ 10 kG and hence the corresponding term can be easily neglected in Eq.~(\ref{rigid5}).  The fast magneto-Rossby wave frequency (with $n<10$) also easily satisfies the weak field condition (Eq.~\ref{ap}) for a field strength $<$ 100 kG, and therefore the fast magneto-Rossby waves do not have a critical layer in Eq.~(\ref{rigid1}).
On the other hand, the slow magneto-Rossby mode frequency yields the dispersion relation
\begin{equation}\label{slow}
{{ \omega}\over {\Omega_0}} \approx {{m \beta^2}\over {2}},
\end{equation}
and ${{ \omega}/{\Omega_0}}$ becomes very small for $\beta \ll 1$. Therefore, $\gamma^2$ is now larger than 1 and the  small $\gamma$ approximation breaks down for the slow mode. The slow magneto-Rossby waves match with the critical layer in Eq.~(\ref{rigid1}) when $m\beta/2 \sim \mu $ and therefore can be unstable. However, periods of slow magneto-Rossby waves are much longer than 1 yr  \citep{Zaqarashvili2015}, therefore the waves are beyond the scope of the paper. \citet{Marquez2017} found unstable magneto-Rossby waves for very strong uniform magnetic field. The possible instability of slow magneto-Rossby waves in the case of the magnetic field profile used here may have similar consequences as in the paper of \citet{Marquez2017}, therefore it is an interesting problem to study in the future.

For the stronger magnetic field strength, the fast and slow modes have the similar frequencies
\begin{equation}\label{fast-slow}
{{ \omega}\over {\Omega_0}} \approx \pm {{m \beta}\over {\sqrt {n(n+1)}}},
\end{equation}
which results in the higher value of $\gamma^2$. Therefore, $\gamma^2 \ll 1$  approximation is valid only for the fast magneto-Rossby mode in the case of weak magnetic field.


In order to find the effect of  $\gamma^2$ on the dispersion relation for fast magneto-Rossby waves, we keep the corresponding term in Eq.~(\ref{rigid5}) and arrive at the dispersion relation
$$
n(n+1) \left ({{ \omega}\over {\Omega_0}}\right )^3  +2m\left ({{ \omega}\over {\Omega_0}}\right )^2-\left ({{9}\over {2}} -{{7}\over {2}} {{4m^2-1}\over {(2n-1)(2n+3)}}\right )m^2\beta^2{{ \omega}\over {\Omega_0}}+
$$
\begin{equation}\label{rigid7}
+m\left (1- {{4m^2-1}\over {(2n-1)(2n+3)}}\right )m^2\beta^2=0.
\end{equation}
Using a transformation of variables in the form of
\begin{equation}\label{rigid8}
{{ \omega}\over {\Omega_0}} ={\tilde \omega} - {{2m}\over {3n(n+1)}}
\end{equation}
one can get a  cubic equation
\begin{equation}\label{rigid9}
{\tilde \omega}^3+p{\tilde \omega} +q=0,
\end{equation}
where
\begin{equation}\label{rigid10}
p=-{{4m^2}\over {3n^2(n+1)^2}} -{{m^2\beta^2}\over {n(n+1)}}\left ({{9}\over {2}} -{{7}\over {2}} {{4m^2-1}\over {(2n-1)(2n+3)}}\right )
\end{equation}
and
\begin{equation}\label{rigid11}
q={{16m^3}\over {27n^3(n+1)^3}}+{{m^3\beta^2}\over {n(n+1)}}\left (1- {{4m^2-1}\over {(2n-1)(2n+3)}}\right )+{{2m^3\beta^2}\over {3n^2(n+1)^2}}\left ({{9}\over {2}} -{{7}\over {2}} {{4m^2-1}\over {(2n-1)(2n+3)}}\right ).
\end{equation}
Inspection of this equation shows that $p<0$ and $4p^3+27q^2>0$ for the  weak magnetic field limit $\beta \ll 1$, therefore it has only one real solution
\begin{equation}\label{rigid12}
{\tilde \omega}=  - 2 {{|q |}\over {q}} \sqrt{- {{p}\over {3}} }\cosh \left ({{1}\over {3}}{\arccosh} \left ( {{-3 |q |}\over {2p}} \sqrt{- {{3}\over {p}} } \right ) \right ).
\end{equation}

%


This is the dispersion relation of fast magneto-Rossby waves and for small $\beta \ll1$ it can be expressed as
\begin{equation}\label{rigid13}
{{ \omega}\over {\Omega_0}} = - {{2m}\over {3n(n+1)}} \left (1+2\sqrt{1+{{81}\over {8}} n(n+1)\beta^2} \right ).
\end{equation}

Figure 3 shows the periods of various magneto-Rossby wave harmonics vs magnetic field strength calculated from Eq.~(\ref{rigid12}). Here the magnetic field strength denotes the maximal value along the latitude, which equals to $B_0/2$. We see that lower order harmonics with $n$ lead to the periodicity of $< 300$ days. Only $n=5$ and $n=6$ harmonics may result in annual oscillations for small value of magnetic field strength.


For small $\gamma^2$, spheroidal wave functions can be expanded in terms of associated Legendre polynomials \citep{abramowitz}
$$
H(\mu)=S_{mn}(\gamma,\mu)=P^m_ {n}(\mu)
$$
\begin{equation}\label{rigid14}
+\left (-{{(m+n-1)(m+n)P^{m}_ {n-2}(\mu)}\over {2(2n-1)^2(2n+1)}} +{{(n-m+1)(n-m+2)P^{m}_ {n+2}(\mu)}\over {2(2n+1)(2n+3)^2}}  \right )\gamma^2.
\end{equation}

Figure 4 shows the latitudinal structure of toroidal velocity and magnetic field perturbations for the spherical harmonics of $n=3$, $n=4$, and $n=5$ with $m=1$. $n=3$ and $n=5$
harmonics are antisymmetric with the equator in velocity and symmetric in magnetic field. On the other hand, $n=4$ harmonic is symmetric in velocity and antisymmetric in magnetic field.

\subsection{Magneto-Rossby waves on differentially rotating surface}

Now we consider nonzero latitudinal differential rotation, $\Omega_d \not=0$, in Eqs.~(\ref{B})- (\ref{Omega}). The latitudinal differential rotation in the lower part of the tachocline may lead to
magnetohydrodynamic instabilities in joint action with the toroidal magnetic field \citep{gilman97,Dikpati2001}. In order to find unstable harmonics, we use the general technique of Legendre
polynomial expansion  \citep{longuet68,gilman97,Zaqarashvili2010a}. In this case, we are not restricted with weak magnetic field approximation, hence the field strength may take any value.

\begin{figure}
   \begin{center}
  \includegraphics[width=18.0cm]{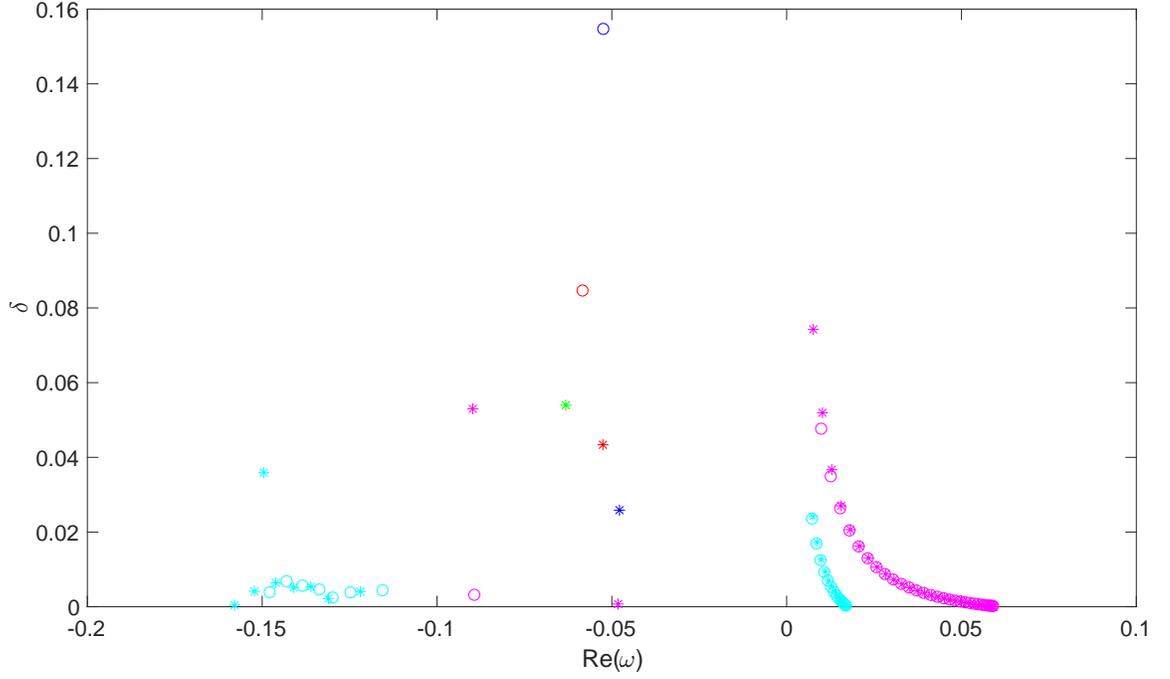}
  \end {center}
  \caption{Normalised growth rates ($\delta=Im(\omega)/Re(\omega)$) vs frequencies of unstable harmonics. Frequencies are in units of  $\Omega_0$. The unstable harmonics with periods of $>$ 10 years are not shown.
Cyan, magenta, green, red and blue asterisks (circles) correspond to the symmetric (antisymmetric) harmonics with the latitudinal peak strength of magnetic field
10 kG, 20 kG, 30 kG, 40 kG, and 50 kG, respectively. Differential rotation parameters are $s_2=s_4=0.13$.}\label{Rossby}
\end{figure}

We expand $\Psi$ and $\Phi$ in infinite series of associated Legendre polynomials
\begin{equation}
 \Psi=\sum^{\infty}_{n=m}a_nP^m_n(\mu),\,\,\,\Phi=\sum^{\infty}_{n=m}b_nP^m_n(\mu),
\end{equation}
substitute them into Eqs.~(\ref{B})- (\ref{Omega}), and using a recurrence relation of Legendre polynomials
we obtain algebraic equations as infinite series \citep{Zaqarashvili2010a}. The dispersion relation for the infinite number of harmonics can be obtained when the infinite determinant of the system is set to zero. In order to solve the determinant, we truncate the series at $n=100$ and solve the resulting polynomial in $\omega$ numerically. Solutions with complex values of $\omega$ are unstable, where $Re(\omega)$ are frequencies and $Im(\omega)$ are growth rates of unstable harmonics.  Frequencies and growth rates depend on the differential rotation parameters, $s_2$ and $s_4$, whose values can be estimated by helioseismology, though with significant uncertainty \citep{Schou1998}. The estimated parameters are close to the  photospheric values at the upper tachocline but become very small near the bottom. We take their values as $s_2=s_4=0.13$, which might correspond to the middle of the tachocline. Figure 5 shows all unstable harmonics on the complex $\omega$-plane for different values of magnetic field strength. Weak magnetic field (10 and 20 kG) allows both eastward or retrograde (with negative frequency) and westward or prograde (with positive frequency) propagating unstable modes. Westward propagating modes have very low frequencies (less than 0.05 $\Omega_0$), while the eastward propagating waves have moderate values between 0.1-0.15 $\Omega_0$.
 The eastward propagating waves are probably connected with the critical point, $\omega_1 =\Omega_d$, in Eqs.~(\ref{B})- (\ref{Omega}), which is due to the differential rotation.
Then the minus sign in $\Omega_d$ requires the minus sign in $\omega_1$, therefore the unstable modes are fast magneto-Rossby waves. On the other hand, westward propagating waves may correspond to the critical point,  $\omega_1 =\beta \mu$, in Eq.~(\ref{rigid1}), which is due to the nonuniform magnetic field. These waves are probably slow magneto-Rossby waves as their frequency is positive (Eq.~\ref{slow}).
Stronger magnetic field ($\geq$ 30 G) removes westward propagating modes and only eastward propagating modes remain whose frequencies tend to 0.05 $\Omega_0$ (with the period of 560 days).  Figure 6 shows the periods of most unstable harmonics vs peak magnetic field strength. Weak magnetic field yields shorter periods with 150-180 days, which correspond to Rieger-type periodicities, while stronger field gives longer periods with 400-600 days, which are in the range of annual oscillations.

 \begin{figure}
   \begin{center}
  \includegraphics[width=18.0cm]{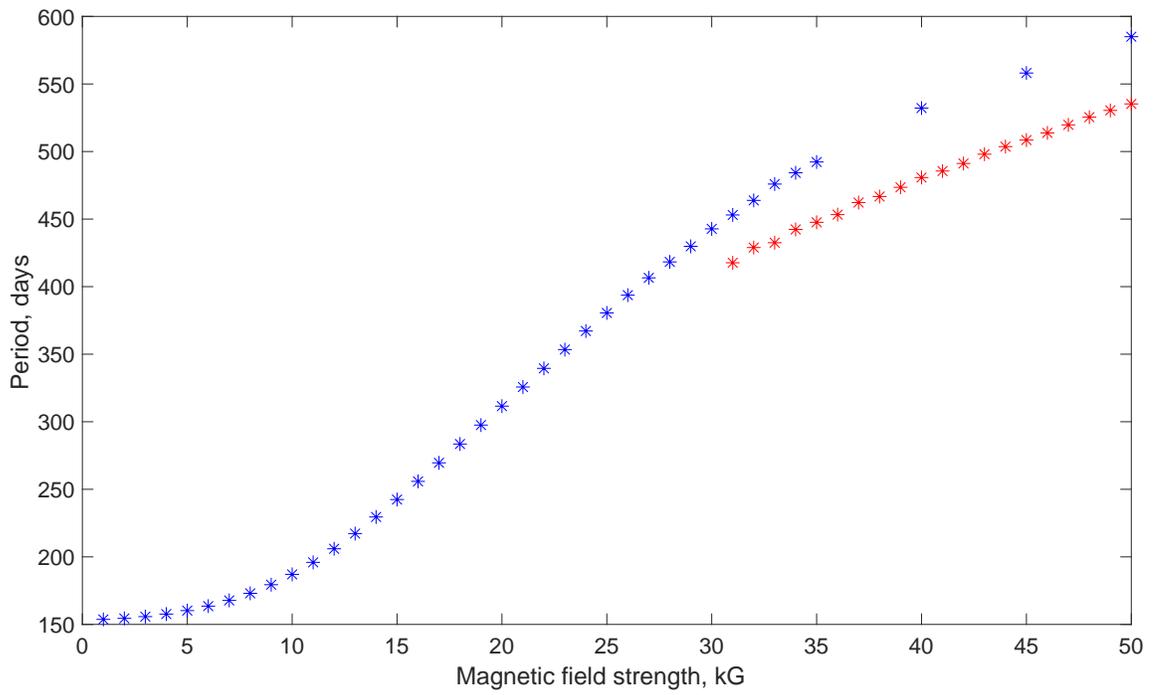}
  \end {center}
  \caption{Periods of most unstable harmonics vs magnetic field strength. Blue (red) asterisks show symmetric (antisymmetric) modes.}\label{Rossby}
\end{figure}

\section{Multiple periodicities in the solar cycle 23}

Theoretical results of previous sections show that different wave modes and different harmonics of the same modes correspond to different time scales. Therefore, it is very  important to search multiple periodicities in solar activity  in the range of 150-500 days. For this reason, we use Greenwich Royal Observatory sunspot area hemispheric data for the solar cycle 23. This cycle displays significant north-south asymmetry, therefore we search  periods in  both hemispheres separately. Cycle 23 was a generally south dominated cycle with two activity maxima, first in the northern hemisphere, and another more powerful maximum in the southern one. The southern hemisphere was dominating duringmost of  the cycle, while the northern hemisphere was dominating only during a short interval of  the ascending phase.

We use Morlet wavelet analysis \citep{To98} to search for the periods in the range of 150-500 days in full disk and hemispheric data. Figure 7 shows the resulting periods which are well seen during solar activity maximum. The shortest period in this range is the well known Rieger type periodicity, which is detected in almost all activity indices \citep{Rieger84,Lean89,carball90,oliver98}. This periodicity has the most remarkable power in both hemispheres (middle and lower panels on the figure), but it is less powerful in full disk data (top panel).  The period is 175 days (160 days) in northern (southern) hemisphere and around 160 days in full disk data.

The next periodicity is located in the interval of 200-300 days. The southern hemisphere shows a significant peak at the period of 270 days, while the northern hemisphere displays less significant peak at the period of 245 days. Full disk data has a peak at 260 days.


Another branch of periodicity is located near the time scale of 310-320 days. This is  a less powerful peak, but it is seen in all three panels of Figure 7. Southern hemisphere shows a peak near 320 days, while northen hemisphere and full disk data display the periodicity of 310 days. This time scale is very close to the period of 323 days reported by  \citet{Lean89} and \citet{oliver92} in the older cycles.

 The next powerful peak is at 380 days in full disk, northern and southern hemispheric data. This peak corresponds to recently reported annual oscillations in coronal bright points \citep{McIntosh2015,McIntosh2017}. Hence, sunspot areas and bright points show the similar periodicity, which may indicate to the same underlying  mechanism.

A very interesting small peak is seen near 460 days in the southern hemisphere, while the similar peak is absent in the northern hemisphere. This peak corresponds to the periodicity of 1.3 years, which has been reported by helioseismology near the base of convection zone in the same cycle 23 \citep{howe00}. This periodicity is quite  an enigmatic phenomenon, it appears in some cycles and disappears in another cycles. \citet{Krivo02} analyzed different sunspot data and reported that the power at the 1.3-year periodicity fluctuates considerably with time, being stronger during stronger sunspot cycles. This periodicity is obviously seen in full disk data in upper panel of Figure 7 as well, but it has very small power. All global wavelet periods are shown on the Table 1.

\begin{figure}
\includegraphics[width=12.0cm]{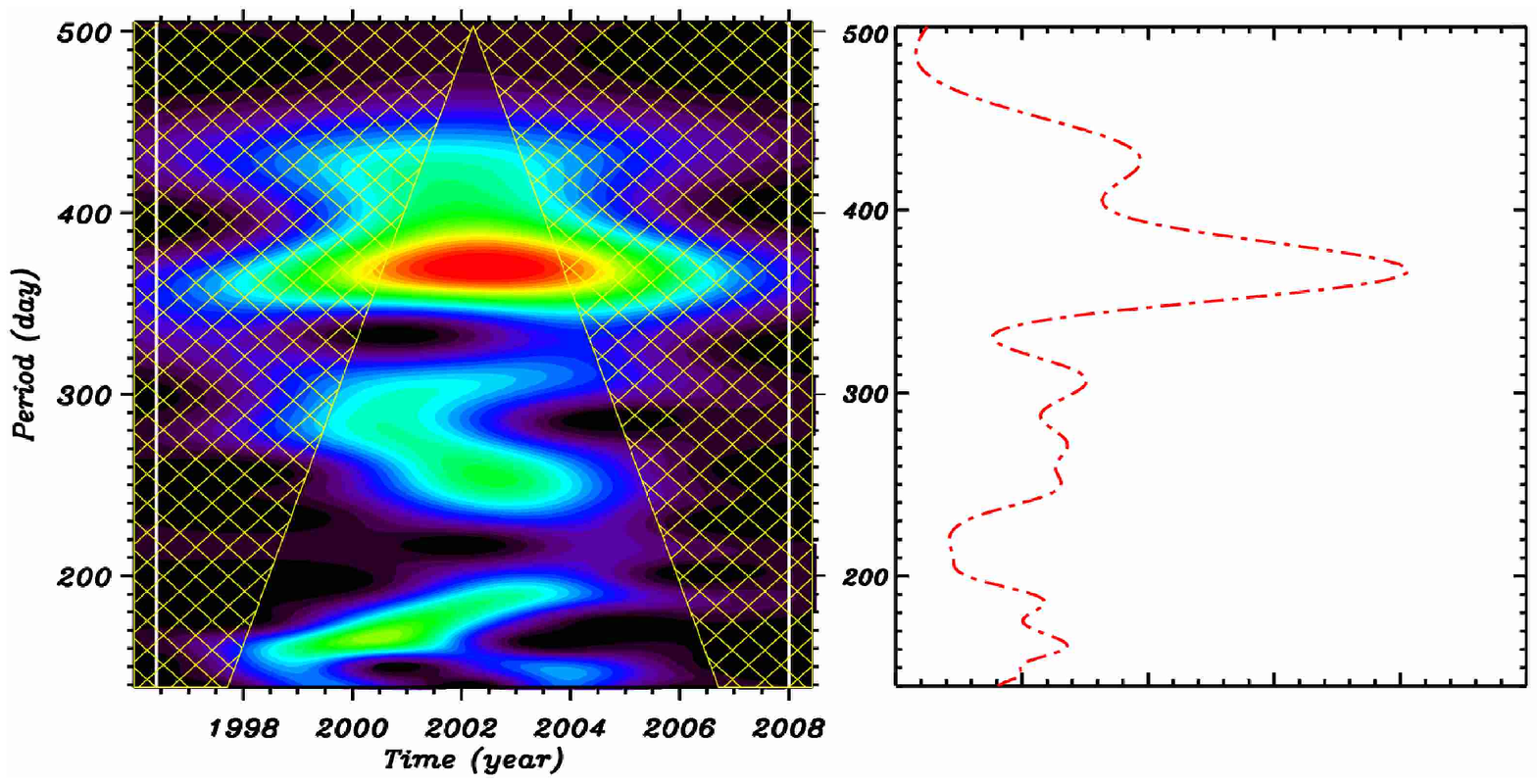}{\centering}
\includegraphics[width=12.0cm]{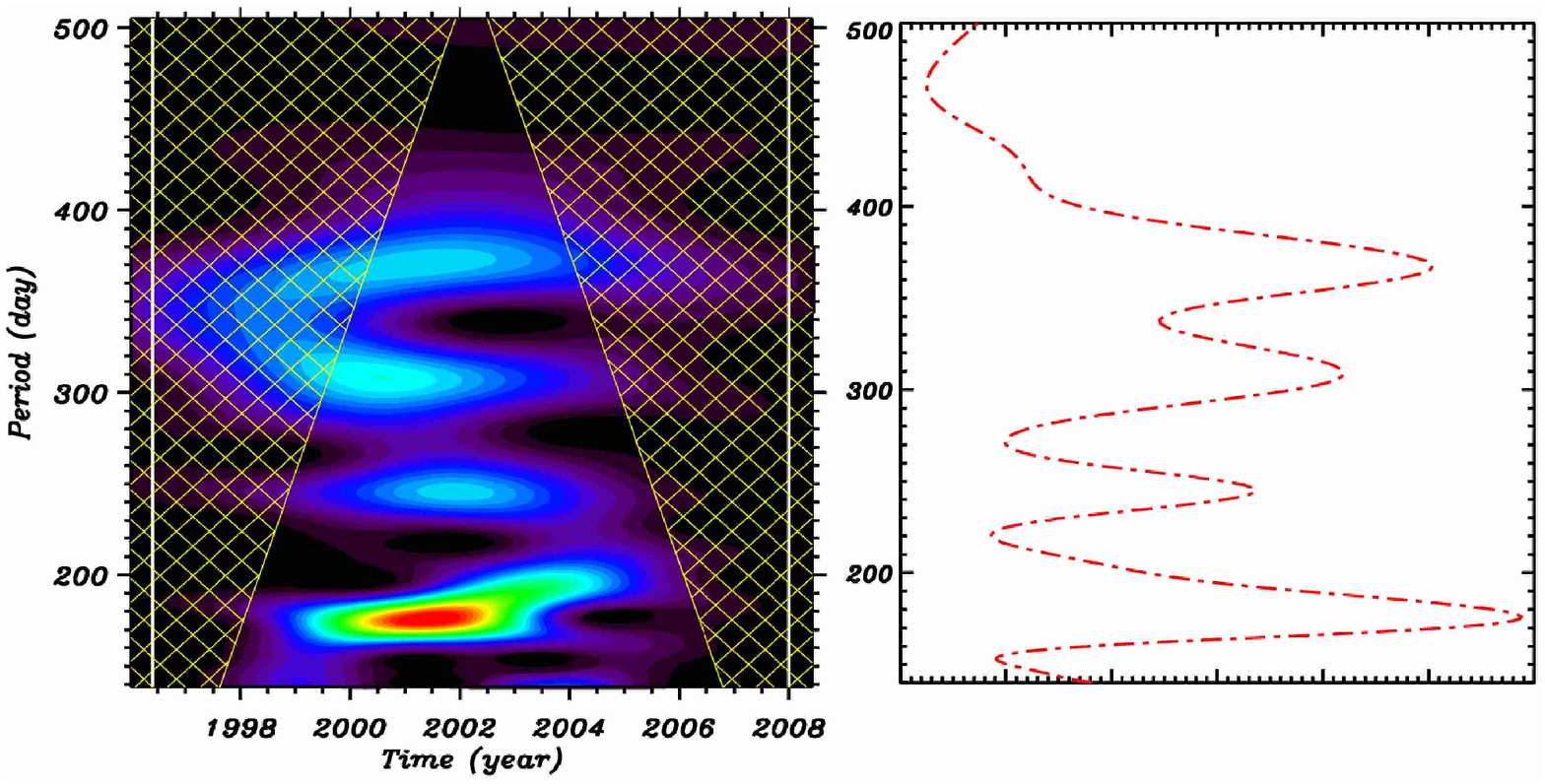}{\centering}
\includegraphics[width=12.0cm]{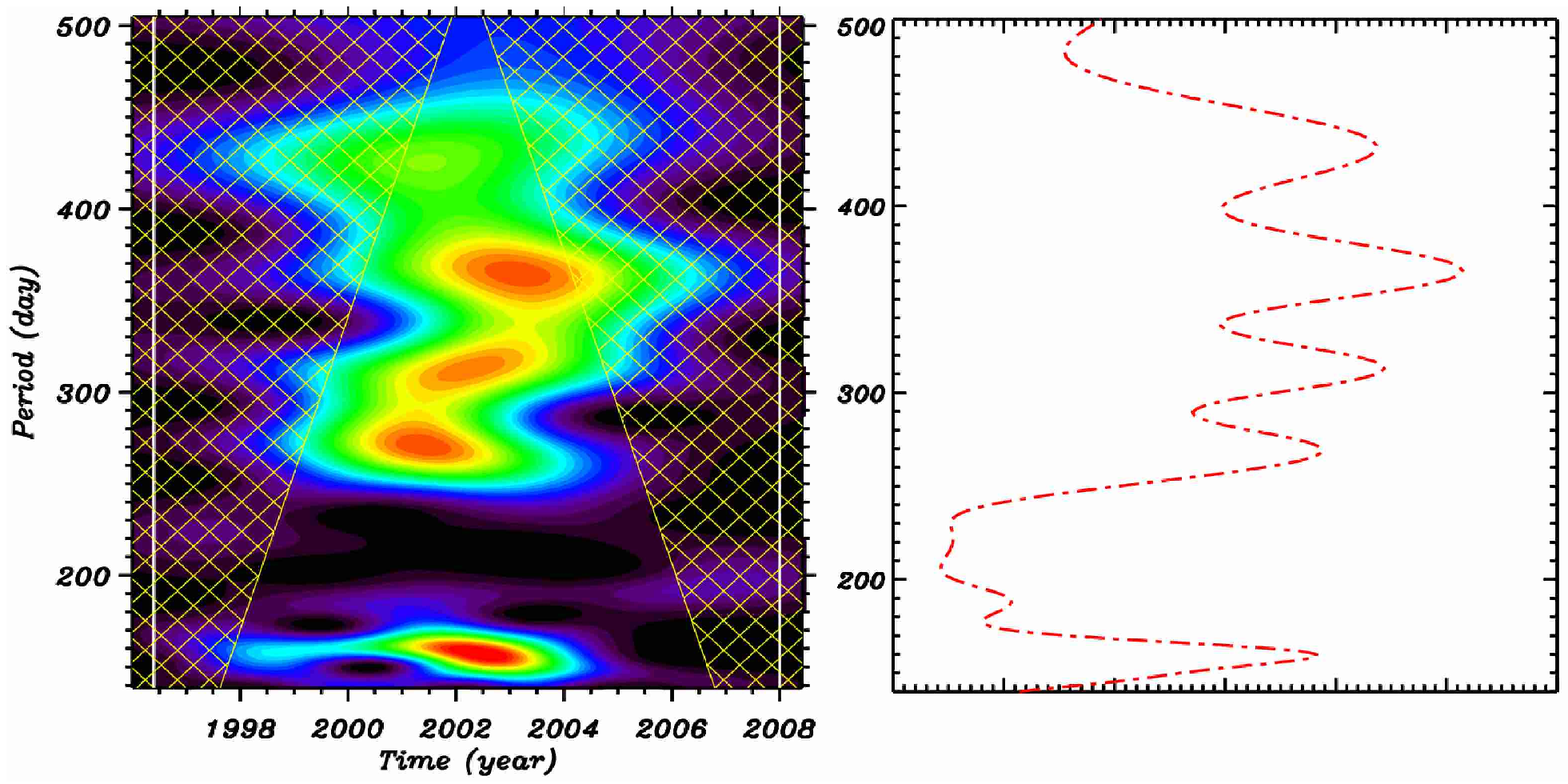}{\centering}
\caption{Morlet wavelet analysis of GRO daily sunspot data. Top, middle and bottom panels correspond to full disk, northern and southern hemispheres, respectively. In hatched areas outside the COI the wavelet transform is not reliable. The global wavelets are computed and plotted alongside each wavelet, where the most powerful peaks are denoted. }
\end{figure}

\begin{table}[h!]
\centering
 \begin{tabular}{||c c c c c c  ||}
  \hline
  & Period (days)  & Period (days) & Period (days) & Period (days) &  Period  (days)  \\
 \hline\hline
 Full disk & 160 & 260 & 310 & 380 & 450 \\
 \hline
 North & 175 & 245 & 310 & 380 & -  \\
 \hline
 South & 160 & 270 & 320 & 380 & 460 \\
 \hline

\end{tabular}
\caption{Estimated mid-range periodicities for the solar cycle 23 using GRO sunspot area hemispheric data. The first, second, third, fourth and fifth columns show most powerful periods in the range of 150-200 days, 200-300 days, 300-350 days, 350-400 days and 400-500 days, respectively. The periods are found by global wavelet analysis from Fig. 7.  }
\label{table:1}
\end{table}

\section{Discussion}

Quasi-annual oscillations in solar activity with a period of 323 days were observed in many activity indices more than twenty years ago \citep{Lean89,oliver92}. Then helioseismic measurements detected the periodicity of 1.3 yrs (around 470 days) in solar differential rotation near the base of convection zone \citep{howe00}. However, no clear physical mechanism for the periodicity has been suggested since the observations. Recently, \citet{McIntosh2017} found the period of 1 year in coronal bright points using joint STEREO and SDO observations. They estimated retrograde phase speed of $\sim$ 3 m s$^{-1}$ with slight difference in the northern and the southern hemispheres. The estimated phase speed is close to the slow magneto-Rossby wave phase velocity \citep{Zaqarashvili2015} in the solar tachocline, therefore \citet{McIntosh2017} interpreted the observations in terms of magneto-Rossby waves.  However, \citet{McIntosh2017} detected  a retrograde phase speed of the wave, which rather corresponds to fast magneto-Rossby waves than to slow modes. But, the periodicity of 1 yr is indeed close to the period of the global unstable magneto-Rossby waves in the solar tachocline \citep{Zaqarashvili2010b}, therefore Rossby-type phenomenon is clearly involved in observed process. Recently, \citet{dikpati2017,dikpati2018} suggested that the periodicity can be caused by nonlinear periodic energy exchange between Rossby waves and differential rotation, where the period shows the time scale of energy exchange. \citet{Zaqarashvili2018} suggested that the annual oscillations could be also explained in terms of equatorially trapped magneto-Kelvin waves in the solar tachocline. Therefore, both Rossby and Kelvin waves might be considered as mechanisms for annual oscillations. On the other hand, wavelet analysis of GRO sunspot area data for the cycle 23 shows multiple periodicity in the interval of 150-500 days for northern and southern hemispheres (see Figure 7 and Table 1). Hemispheric and full disk periodicities are found in the range of 450-460, 370-380, 310-320, 240-270, and 150-175 days. The different periodicities may correspond to different wave modes or different harmonics of the same mode. The only measured toroidal phase speed is for 1 yr oscillations \citep{McIntosh2017}, therefore it is much easier to determine which wave mode is responsible for this oscillation. This periodicity is seen in GRO sunspot area data as 380 days by wavelet analysis (see Table 1). We discuss a possible role of magneto-Kelvin and magneto-Rossby waves in excitation of the periodicity.

Global magneto-Kelvin waves with spatial scale of equatorial extend i. e. $m=k_xR=1$, where $m$ ($k_x$) is a toroidal wave number in spherical (rectangular) coordinates, may have the period of 1 yr for particular values of normalised reduced gravity (see Fig. 1) in the upper overshoot tachocline. The solutions of magneto-Kelvin waves are located between $\pm 40$ degree in the case of nonuniform toroidal magnetic field $B_0\sin{\hat \theta} \cos{\hat \theta}$, where ${\hat \theta}$ is the latitude. Dispersion relation of magneto-Kelvin waves (Eq.~\ref{kelvin21}) gives the phase speed of the waves as $v_{ph}=c=\sqrt{gH}$. If one uses the normalised reduced gravity of $G=0.006$, which actually gives 1 yr oscillations for $m=1$ harmonic (Fig. 1), then the phase speed is around $\sim$ 100 m s$^{-1}$. This is much higher than the observed speed of 3 m s$^{-1}$. Another problem is that the phase speed of magneto-Kelvin waves is prograde  for weak/moderate magnetic field strength, i.e. the waves propagate in the direction of rotation, while the observed pattern is retrograde \citep{McIntosh2017}.  Very strong magnetic field ($>$ 200 kG) lead to a retrograde pattern for magneto-Kelvin waves (see also \citet{Marquez2017}), but this possibility seems unlikely in our case for two reasons. First, the field seems to be stronger than it is believed to be in the tachocline. Second, the phase speed of magneto-Kelvin waves in the case of such strong magnetic field reaches 2 km s$^{-1}$, which is much higher than the observed value. Therefore, magneto-Kelvin waves do not match with observations. On the other hand, the global magneto-Kelvin waves with m=1 and $G=0.004$ (Fig. 1) might be responsible for 1.3 yrs oscillations (the fifth column on Table 1).

\begin{figure}
\includegraphics[width=15.0cm]{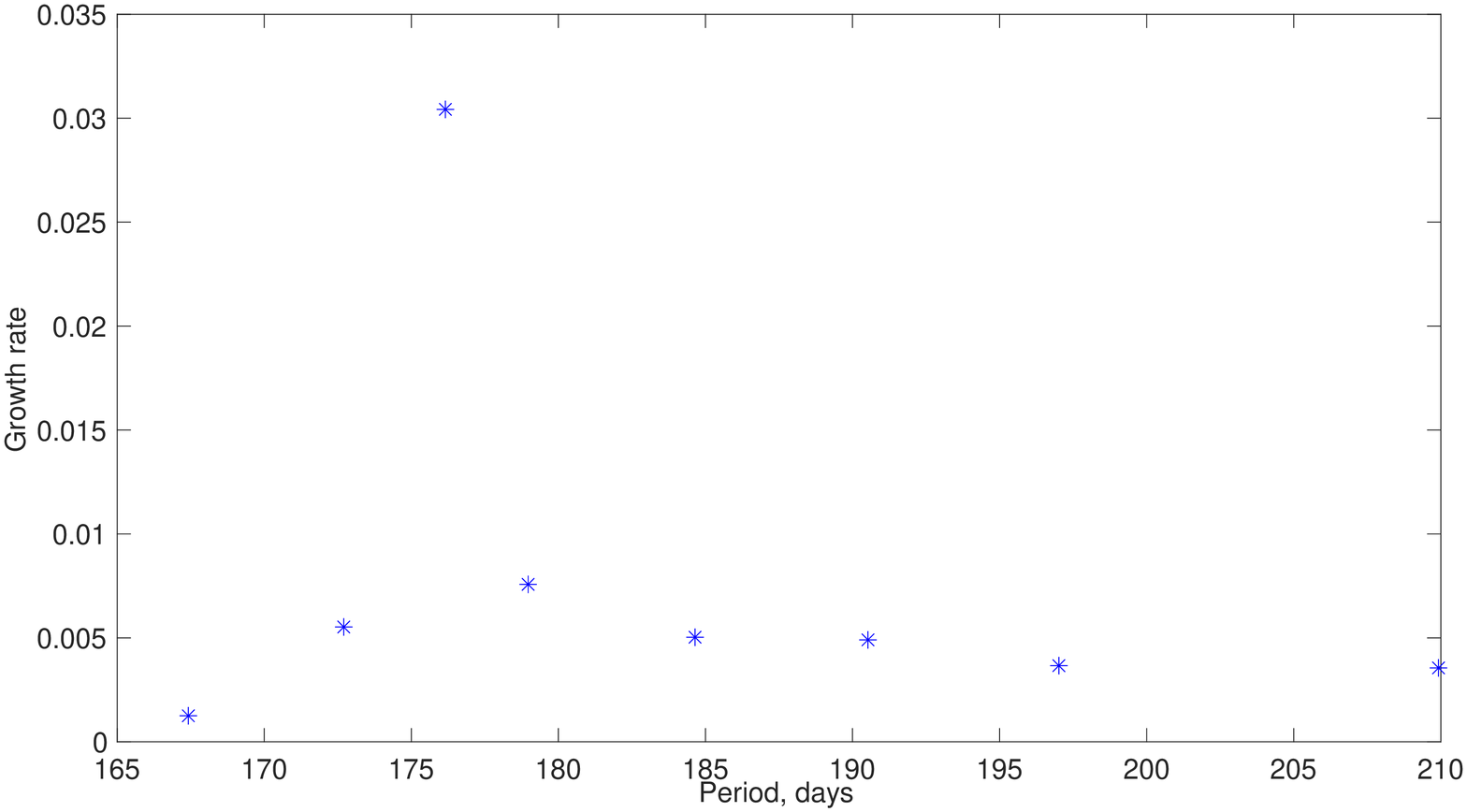}{\centering}
\includegraphics[width=15.0cm]{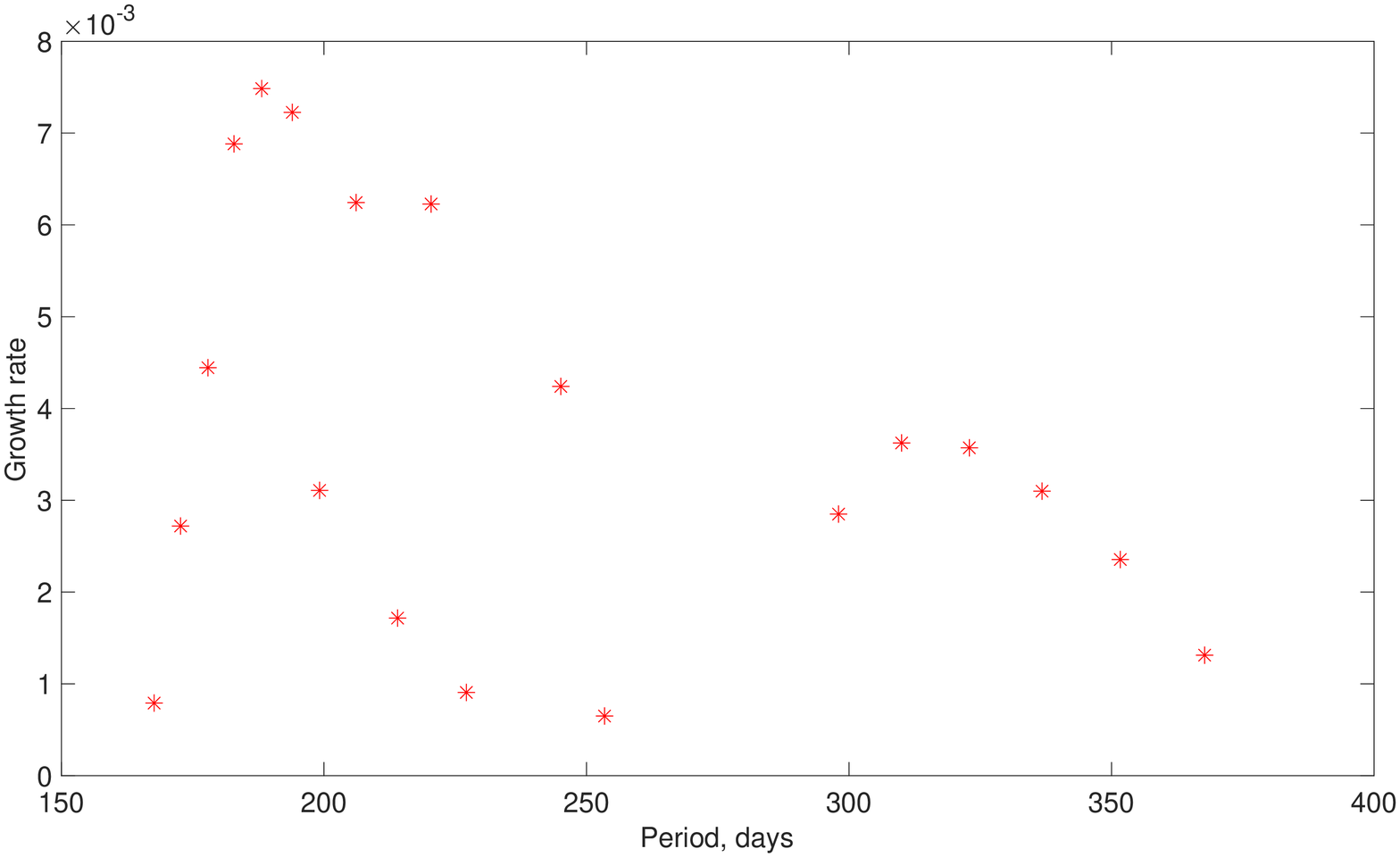}{\centering}
\caption{Normalised growth rates ($\delta=Im(\omega)/Re(\omega)$) vs periods of unstable magneto-Rossby wave harmonics for the magnetic field strength of 7 kG and the differential rotation parameters of $s_2=s_4=0.12$. Upper panel (blue asterisks) shows symmetric modes. Lower panel (red asterisks) shows antisymmetric modes. }
\end{figure}

Global magneto-Rossby waves may also lead to an observed periodicity of 1 yr  for $\epsilon \ll 1$.  In this case, one can consider spherical surface instead of shallow layer, which significantly simplifies the calculations.  Analytical solutions show that the spherical harmonics of fast magneto-Rossby waves with $m=1$ and $n=5$ (or $n=6$) have the time scale of one year, hence can be responsible for the observed periodicity. Fast magneto-Rossby waves are retrograde, hence they correspond to the observed pattern. On the other hand, the phase speed of $m=1$ and $n=6$ harmonic  in the case of 6 kG magnetic field, which gives 1 yr period without differential rotation (see subsection 4.1), can be estimated as 67.8 m s$^{-1}$ from Eq.~\ref{rigid13} (phase speeds for $n=6, 5, 4, 3$ are shown on Table 2.). The speed is much higher than the observed speed of 3 m s$^{-1}$ \citep{McIntosh2017}. However, it must be noted that the phase speed of 67.8 m s$^{-1}$ is estimated with regards to the equatorial angular velocity, $\Omega_0$. On the other hand, at higher latitudes global Rossby wave pattern must show lower apparent phase speed with regards to the local surface because of solar differential rotation (higher latitudes rotate slower than the equator). For example, the solar surface at the latitude 40$^0$ moves with the speed of 65 m s$^{-1}$  opposite to the equatorial rotation if the differential rotation rate is $s_2=0.2$. Hence the global Rossby wave pattern moving with the phase speed of 67.8 m s$^{-1}$ with regards the equator will show the apparent phase speed of 2.8 m s$^{-1}$ in Hovm\"oller diagrams at the latitude 40$^0$. This is very close to the observed value by \citet{McIntosh2017}. Hence, the global magneto-Rossby wave with $m=1$ and $n=6$ might be responsible for the oscillations of 380 day periodicity (see Table 1) for the magnetic field strength of 6-7 kG. For the same magnetic field strength $m=1$ and $n=5$ harmonic gives the period of 310-320 days (see Fig. 3), which corresponds to  the next observed branch on Table 1. The $n=4$ harmonic gives the periodicity of 240-270 days matching with  the next period branch on the Table 1. Finally,  the $n=3$ harmonic gives the Rieger-type periodicity of 150-170 days corresponding  to the last shortest period from Table 1. Therefore,  fast global magneto-Rossby wave harmonics with $n=6, 5, 4, 3$ and $m=1$ may explain the observed multiple periodicities in GRO sunspot area during cycle 23. However, the analytical solutions which match with the observed periodicity are valid only for weak field approximation (several kG). When the magnetic field strength increases, the dispersion relation does not describe the estimated periods properly, therefore the result must be treated with caution.

\begin{table}[h!]
\centering
\begin{tabular}{||c|c|c|c|c||}
  \hline
  n & 6 & 5 & 4 & 3 \\
    \hline
  Phase speed & 67 m s$^{-1}$ & 94 m s$^{-1}$ & 137 m s$^{-1}$ & 224 m s$^{-1}$ \\
  \hline
\end{tabular}
\caption{Phase speeds of fast magneto-Rossby waves for  $n=6, 5, 4, 3$ and $m=1$  in the case of 6 kG magnetic field.}
\label{table:1}
\end{table}

Latitudinal differential rotation together with the toroidal magnetic field leads to the instability of magneto-Rossby wave harmonics. Figure 5 shows that the fixed parameters of magnetic field and differential rotation favour one or two most unstable harmonics of retrograde modes (with negative frequency). For the weak field, the prograde harmonics (with positive frequency) are also unstable, but they disappear for the stronger field.  The retrograde waves are fast magneto-Rossby waves and the prograde waves are slow magneto-Rossby waves. The interesting point is that the weak magnetic field provokes many unstable harmonics, while the stronger field allows to occur only one or two modes. The period of  the most unstable harmonic is near the Rieger time scale for weaker field strength (less than 10 kG), but increases up to the time scale of 1-2 years for stronger field of 30-50 kG (Figure 6). Therefore, the global unstable harmonics of magneto-Rossby waves could be responsible for the observed multiple periodicity only for weak magnetic field ($<$ 20 kG). In order to compare the analytical results to unstable modes, we calculated the unstable spectrum for the same magnetic field strength of 7 kG (see previous paragraph). Figure 8 shows the growth rates of all unstable harmonics of retrograde magneto-Rossby waves with $m=1$ for 7 kG toroidal field and for the differential rotation parameters $s_2=s_4=0.12$.  The symmetric harmonic spectrum (upper panel) is dominated by the unstable harmonic with a period of 175 days, which is in the range of Rieger-type periodicity. On the other hand, antisymmetric spectrum (lower panel) can be formally divided into two branches of unstable harmonics. One branch has  periods of $<260$ days, while the other branch has periods of  $> 300$ days. The first branch can be responsible for the observed periods of 150-170 and 240-270 days, while the second branch may account for the observed periods of 310-320 and 380 days. These simple calculations slow that the Rossby wave harmonics may, in principle, explain the observed multiple periodicity, but more sophisticated consideration is necessary.

It must be also noted that some of observed periods might be related to the time scale of energy exchange between differential rotation and unstable Rossby waves as suggested by \citet{dikpati2017,dikpati2018}. These authors concluded that the time scale is of the order of 1 yr. But obviously more analytical/numerical study should be performed in order to make a firm conclusion.

\section{Conclusions}

Linear analysis of global spherical wave modes in the solar tachocline shows that the magneto-Kelvin and  fast magneto-Rossby waves may give  multiple periodicity in the range of 150-500 days. Magneto-Kelvin waves propagate in the direction of solar rotation (for moderate value of magnetic field strength) and fast magneto-Rossby waves propagate in the opposite direction. Therefore, the retrograde phase speed of 1-yr oscillations observed by \citet{McIntosh2017} could be explained in terms of  fast magneto-Rossby waves rather than magneto-Kelvin waves. Morlet wavelet analysis of GRO data for the cycle 23 showed  multiple periodicity of 450-460, 370-380, 310-320, 240-270, and 150-175 days. A longer periodicity of 450-460 days is probably related to the observed 1.3 yr oscillations  \citep{howe00} and might be explained by magneto-kelvin waves. The other periodicities could be explained by different spherical harmonics of  fast magneto-Rossby waves with $m=1$ and $n=6, 5, 4, 3$ in the case of weak toroidal magnetic field with the strength of $<$ 10 kG.  It must be noted that very slow retrograde phase speed observed by \citet{McIntosh2017} could be explained by the latitudinal differential rotation, which actually slows down  the apparent retrograde phase speed of magneto-Rossby waves with regards to the local surface rotation.  More sophisticated analysis (also with more complex numerical simulations) may lead to the development of tachocline seismology examining by long-term periodicities in solar activity.

\acknowledgments

This work was supported by the Austrian Science Fund (FWF) project P30695-N27 and by Georgian Shota Rustaveli National Science Foundation project 217146. This paper is resulted from discussions at the workshop of ISSI (International Space Science Institute) team (ID 389) "Rossby waves in astrophysics" organized in Bern (Switzerland). M.C.,R.O. and J.L.B. acknowledge financial support from MINECO under grants AYA2014-54485-P and AYA2017-85465-P and FEDER Funds.
 The authors are indebted to the referee for making comments that have helped clarify some important issues of this paper.

\end{document}